\documentstyle[aps,preprint,floats,epsfig]{revtex}

\begin{document}

\tightenlines
\hfuzz=4pt

\preprint{
\font\fortssbx=cmssbx10 scaled \magstep2
\hbox to \hsize{
{\fortssbx University of Wisconsin - Madison}
\hfill$\vcenter{\tighten
		\hbox{\bf MADPH-01-1233}
                \hbox{\bf IFT--P.048/2001}
                \hbox{July 2001}}$}}

\title{\vspace*{.5in}  
Soft Color Enhancement of the Production of $J/\psi$'s by Neutrinos}

\author{O.\ J.\ P.\ \'Eboli$^{1}$, E.\ M.\ Gregores$^{2}$, and F.\ Halzen$^3$}

\address{\vspace*{2mm}
$^1$Instituto de F\'{\i}sica, Universidade de S\~ao Paulo,
S\~ao Paulo -- SP, Brazil. \\
$^2$Instituto de F\'{\i}sica Te\'orica, Universidade Estadual Paulista,
S\~ao Paulo -- SP, Brazil. \\
$^3$Department of Physics, University of Wisconsin,
Madison -- WI, USA.
}

\maketitle

\thispagestyle{empty}

\begin{abstract}

\vskip-5ex 

We calculate the production of $J/\psi$ mesons by neutrino--nucleon
collisions in fixed target experiments. Soft color, often referred to as
color evaporation effects, enhance production cross sections due to the
contribution of color octet states. Though still small, $J/\psi$ production
may be observable in present and future experiments like NuTeV and $\mu$
colliders.

\end{abstract}

\draft

\newpage

\section{Introduction}

The power of neutrino beams for probing the structure of the nucleon as
well as general properties of QCD, has been demonstrated by many
experiments \cite{LlewellynSmith:1972zm,Conrad:1998ne,Yang:2000ju}.  We
revisit here the measurement of charm production in neutrino--nucleon
interactions, which has been extensively studied
\cite{Abramowicz:1982wc,Astier:2000us,Adams:2000mn,Alton:2000ze}. 
Because of its striking experimental signature into dimuons, the
neutral current production of charm as a $J/\psi$ bound state is of
particular interest \cite{Barger:1980kb}. This process only happens at
an observable rate provided color octet states also lead to the
formation of charmonium in addition to the traditional color singlet
contributions \cite{csm}.

It is now clear that $J/\psi$ production is a two-step process where a
heavy quark pair is produced first, followed by the non-perturbative
formation of the colorless asymptotic state. As a consequence, color
octet as well as singlet $c\bar{c}$ states contribute to the production
of $J/\psi$. This is clearly supported by the data\cite{cdf,d0,com,Amundson:1997qr}. Two formalisms have been proposed
to incorporate these features: the Non-Relativistic QCD (NRQCD)
\cite{nrqcd}, and the Soft Color (SC) scheme
\cite{cem,Amundson:1996em}.  Experiments measuring the polarization of
bound charm and beauty mesons, or, more precisely, the absence of it,
now clearly favor the second framework.

We therefore reevaluate here the production of $J/\psi$ by neutrino
beams in the SC scheme; for comparison, see Ref.~\cite{Petrov:1999fm}
which contains the NRQCD results. The basic SC assumption is that no
observable dynamics is associated with the soft processes that connect
the color of the perturbative produced charm pair with the colorless
charmonium bound state. This scheme, although far more restrictive than
other proposals, successfully accommodates all features of charmonium
and bottomonium production.  Earlier quarkonium production computations
already referred to it as the color evaporation model\cite{cem}.  It
correctly predicts the energy and final state momentum dependence of
charmonium and bottomonium hadro- and photoproduction at all energies,
as well as their production in electron-positron colliders.  This
approach to color is also used to formulate a successful prescription
for the production of rapidity gaps between jets at Tevatron
\cite{faro,D0gap,Eboli:2000dd} and HERA \cite{Eboli:2000dd,buch}.

The SC formalism predicts that the sum of the cross sections of all
onium and open charm states is described by \cite{Amundson:1996em,faro}
\begin{equation}
\sigma_{\rm onium} = \frac{1}{9} \int_{2 m_c}^{2 m_D} dM_{c \bar{c}}~
\frac{d \sigma_{c \bar{c}}}{dM_{c \bar{c}}} \; ,
\label{sig:on}
\end{equation}
and
\begin{eqnarray}
\sigma_{\rm open} &=& \frac{8}{9}  \int_{2 m_c}^{2 m_D} dM_{c \bar{c}}~
\frac{d \sigma_{c \bar{c}}}{d M_{c \bar{c}}}
+ \int_{2 m_D} d M_{c \bar{c}}~\frac{d \sigma_{c \bar{c}}}{dM_{c \bar{c}}}
\; ,
\label{sig:op}
\end{eqnarray}
where $M_{c\bar c}$ is the invariant mass of the $c\bar c$ pair. The
factor 1/9 stands for the probability that a pair of charm quarks
formed at a typical time scale $1/M_\psi$ ends up as a color singlet
state after exchanging an uncountable number of soft gluons with the
reaction remnants. One attractive feature of this model is the above
relation between the production of charmonium and open charm which
allows us to use the open charm data to normalize the perturbative QCD
calculation, and consequently to make more accurate predictions for
charmonium cross sections.

The fraction $\rho_\psi$ of produced onium states that materialize as
$\psi$,
\begin{equation}
\sigma_\psi = \rho_\psi~\sigma_{\rm onium} \; ,
\label{frac}
\end{equation}
has been inferred from low energy measurements to be a constant
\cite{gavai,schuler}. From the charmonium photoproduction, we
determined that $\rho_\psi=0.43$--0.5 \cite{Amundson:1997qr}, and even
this parameter can be accounted for by statistical counting of final
states \cite{ingelman}. The fact that all $\psi$ production data are
described in terms of this single parameter, fixed by $J/\psi$
photoproduction, permitted us to correctly predict a rate for $Z$-boson
decay into prompt $\psi$ \cite{us-z0} an order of magnitude larger then
the color singlet predictions, and to explain the observed production
at the Tevatron \cite{Eboli:1999hh} and HERA \cite{Eboli:1999xx}.
Therefore, let us study the SC prediction for charmonium production in
neutrino--nucleon collisions.


\section{Results}

The NuTeV Collaboration recently reported on the first observation of open
charm production in neutral current deep inelastic neutrino scattering
\cite{Alton:2000ze}. The observed production rate is consistent with a pure
gluon-$Z^0$ boson fusion, and the observed level of charm production was used
to determine the effective charm mass. They found that a value of
\begin{equation}
m_c = 1.40^{+0.83}_{-0.36} \pm 0.26 {\rm GeV}
\end{equation}
best describes the total open charm production for an average 
neutrino energy $\langle E_\nu \rangle = 154$ GeV.

The SC contribution for the prompt $J/\psi$ is directly connected to  open
charm total cross section; see Eqs.~(\ref{sig:on}) and (\ref{sig:op}). The same
value of the naked charm mass that best describes the open charm production
rate must also describe the bound states, as they are both produced through the
same leading perturbative subprocess $\nu+g \to \nu+c\bar{c}$.  To evaluated
the $J/\psi$ production via neutral current we used the package MADGRAPH
\cite{madg} and HELAS routines \cite{helas} to obtain the full-tree level
scattering amplitude. We used the GRV94-LO \cite{grv} parton distribution
function, adjusting the renormalization and factorization scales as appropriate
for a leading order calculation; we choose $\mu_R = \mu_F = Q^2+4m_c^2$,  where
$Q^2$ is the momentum transfered from the leptonic system. The strong coupling
constant was evaluated in leading order with $\Lambda_{QCD}=300$ MeV, and the
fraction of color singlet $c\bar c$ pairs with invariant mass below the open
flavor threshold that hadronizes as $J/\psi$ was assumed to be
$\rho_\psi=0.5$.  Using the central value $m_c=1.40$ GeV, we obtained
\begin{equation}
\sigma(\nu N \to \nu J/\psi X)
= 9.0^{+8.1}_{-5.7}\times 10^{-2} \, {\rm fb}
\end{equation}
for the NuTeV experiment, where the errors reflect the systematic uncertainty
on the charm  mass measurement.

\begin{figure}
\begin{center}
\mbox{\epsfig{file=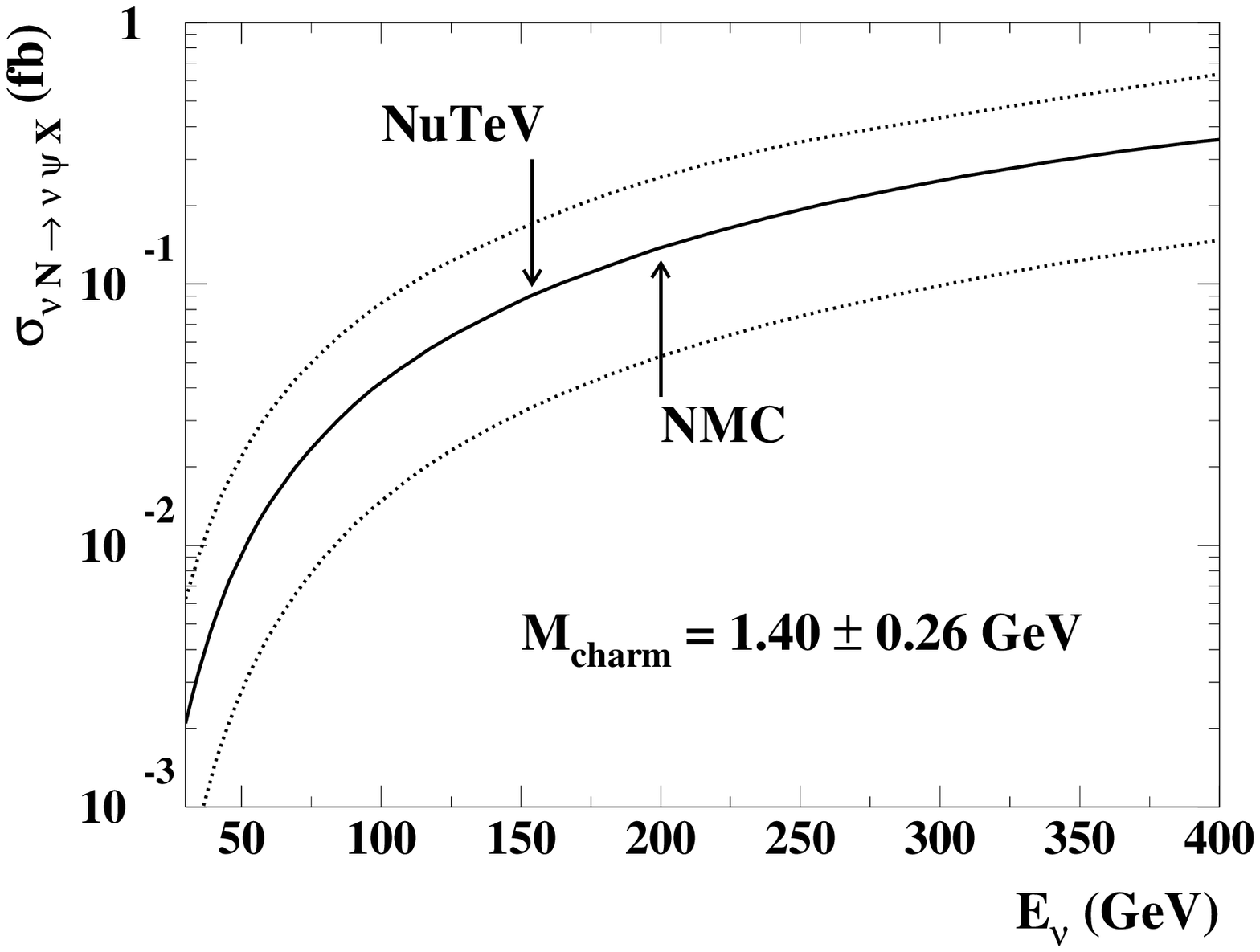,width=0.6\linewidth}}
\end{center}
\caption{Cross section of the reaction $\nu N \to \nu J/\psi X$ as
function of the energy of the incident neutrino beam for three values
of the charm quark mass. The central (upper | lower) curve was obtained using
$m_c=1.40$ ($1.14$ | $1.66$) GeV. The arrows indicate the average beam energy
for the NuTeV experiment and a future muon collider (NMC). }
\label{fig:xsec}
\end{figure}

Using these same choice of parameters we extrapolated the total $J/\psi$
production cross section for other values of $\langle E_\nu \rangle$; the
result is presented in Fig.~\ref{fig:xsec}, where we pinpoint the mean neutrino
beam energy achieved at NuTeV, as well as the value expected at planned muon
collider facilities. In order to show the cross section dependence on the
uncertainty of the charm mass, we also present in this figure the error band
that corresponds to the systematic error on the charm mass as measured by
NuTeV. 

These results clearly indicate that the detection of $J/\psi$'s at NuTeV is a 
challenging task. For instance, taking into account that NuTeV observed about
1.3 million DIS events, with $\sigma_{total}=0.82$ pb, we could expect between
3 and 16 dimuons events to be produced from $J/\psi$ decays. 

The planned muon colliders should be able to generate neutrino beams with
sufficient high luminosity to clearly observe $J/\psi$ events. Moreover, at
these machines the neutrino flux can be accurately calculated
\cite{Harris:1997xi}, allowing more precise predictions.  For instance, a $2
\times 250$ GeV muon collider produces a collimated neutrino beam with average
energy $\langle E_\nu \rangle = 200$ GeV \cite{Harris:1997xi}.  If we assume
the general purpose detector suggested by King \cite{king} with $\simeq$ 50
g/cm$^2$ density, we should have the production of 35--180 dimuons events per
year originated from $J/\psi$ decays, which would allow a more detailed study
of its production properties.

\begin{figure}
\begin{center}
\mbox{\epsfig{file=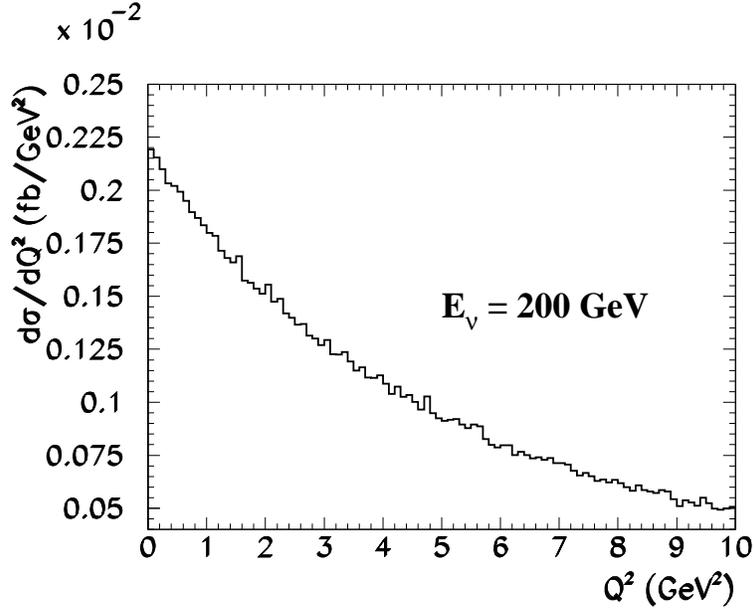,width=0.6\linewidth}}
\end{center}
\caption{Differential cross section of $J/\psi$ as function of the 
squared  momentum transfered by the leptonic system.}
\label{fig:q2}
\end{figure}

\begin{figure}
\begin{center}
\mbox{\epsfig{file=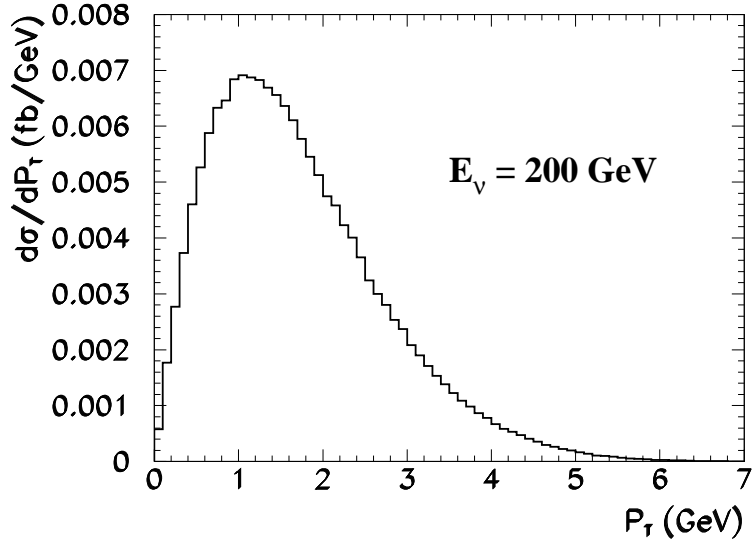,width=0.6\linewidth}}
\end{center}
\caption{Transverse momentum differential cross section of the produced
$J/\psi$'s.}
\label{fig:pt}
\end{figure}

\begin{figure}
\begin{center}
\mbox{\epsfig{file=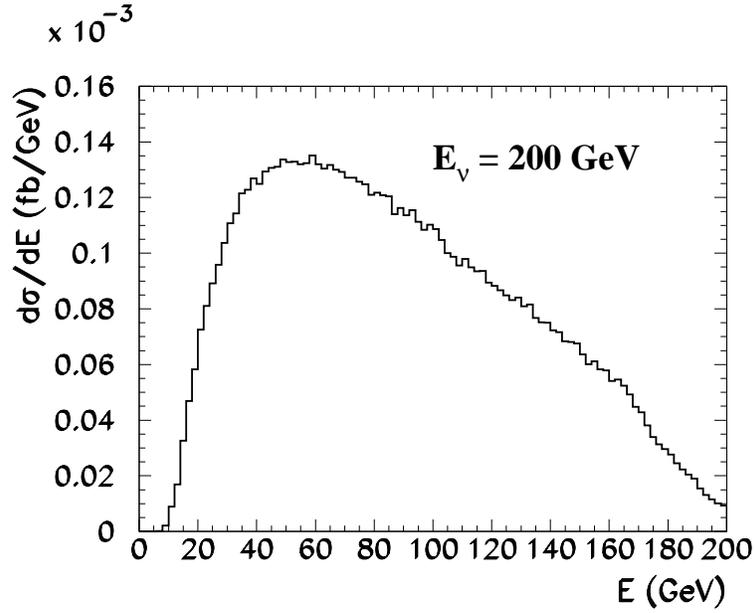,width=0.6\linewidth}}
\end{center}
\caption{$J/\psi$ energy distribution for an incident 
neutrino beam of 200 GeV.}
\label{fig:e}
\end{figure}

In Fig.~\ref{fig:q2} we present the $Q^2$ distribution for a neutrino beam with
average energy $\langle E_\nu \rangle = 200$ GeV \cite{Harris:1997xi}.
Figure~\ref{fig:pt} shows that $J/\psi$'s will be mostly produced with a
small transverse momentum, however the $J/\psi$'s will carry a rather large
energy (on average $88.7$ GeV); see Fig.~\ref{fig:e}.


\section{Conclusion}

Using the NuTeV analysis of open charm production, we calculated the $J/\psi$
cross section in the soft color model.  We found that $\sigma(\nu_\mu N \to
\nu_\mu J/\psi X) = 9.0^{+8.1}_{-5.7}\times 10^{-2} $ fb at the average
neutrino beam energy of the NuTeV experiment. We subsequently calculated
dependence of the $J/\psi$ cross section as a function of the average energy of
the incident neutrino beam, and displayed some differential cross sections for
a neutrino beam associated with a future muon collider.


\acknowledgments

One of us (EMG) would like to thank the University of Wisconsin for its
kind hospitality during the elaboration of this work.  This research
was supported in part by the University of Wisconsin Research Committee
with funds granted by the Wisconsin Alumni Research Foundation, by the
U.S.\ Department of Energy under grant DE-FG02-95ER40896, by
Funda\c{c}\~{a}o de Amparo \`a Pesquisa do Estado de S\~ao Paulo
(FAPESP), by Conselho Nacional de Desenvolvimento Cient\'{\i}fico e
Tecnol\'ogico (CNPq), and by Programa de Apoio a N\'ucleos de
Excel\^encia (PRONEX).


\newpage


\end{document}